\documentclass{aa}  

\usepackage{graphicx}
\usepackage[varg]{txfonts}
\usepackage[T1]{fontenc}
\usepackage{ae,aecompl}

\usepackage{amsmath}	
\usepackage{amssymb}	
\usepackage{times} 
\usepackage{mathptmx}
\usepackage{listings}
\usepackage{gensymb}
\usepackage{amsmath,xspace}
\usepackage{pdflscape}
\usepackage{rotating,booktabs}
 \usepackage{tabularx,tablefootnote} 
\usepackage{natbib}
\bibpunct{(}{)}{;}{a}{}{,} 

\usepackage[usenames]{color} 
\usepackage[normalem]{ulem}
\usepackage[dvipsnames]{xcolor}
\definecolor{green}{rgb}{0.3,0.7,0.}

\let\oldAA\AA
\renewcommand{\AA}{\text{\normalfont\oldAA}}
\newcommand{\rmca}{RMC~127\xspace}
\newcommand{\sixtyone}{S$\,$61\xspace}

\newcommand{\rmc}{RMC$\,$143\xspace}
\newcommand{\dor}{30~Dor\xspace}

\begin{document}

   \title{A massive nebula around the Luminous Blue Variable star \rmc revealed by ALMA}

   \author{C. Agliozzo
          \inst{1,}\inst{2,}\inst{3},
         A. Mehner\inst{1},
         N. M. Phillips\inst{2},
         P. Leto\inst{4},
         J. H. Groh\inst{5},
        A. Noriega-Crespo\inst{6},
        C. Buemi\inst{4},
        F. Cavallaro\inst{4},
        L. Cerrigone\inst{7},
        A. Ingallinera\inst{4},
        R. Paladini\inst{8},
        G. Pignata\inst{3,9},
        C. Trigilio\inst{4},
        G. Umana\inst{4}
          }

   \institute{European Southern Observatory, Alonso de Cordova 3107, Vitacura,  Santiago de Chile, Chile\\
   \email{claudia.agliozzo@eso.org}
         \and
             European Southern Observatory, Karl-Schwarzschild-Strasse 2, Garching bei M\"unchen, 85748, Germany
             \and
             Departamento de Ciencias Fisicas, Universidad Andres Bello,  
            Avda. Republica 252, Santiago, 8320000, Chile
            \and
            INAF-Osservatorio Astrofisico di Catania, Via S. Sofia 78, 95123 Catania, Italy
            \and
            Trinity College Dublin, The University of Dublin, College Green, Dublin, Ireland
            \and
            Space Telescope Science Institute 3700 San Martin Dr., Baltimore, MD, 21218, USA
            \and
            Joint ALMA Observatory, Alonso de C\'{o}rdova 3107, Vitacura, Santiago, Chile
            \and
            Infrared Processing Analysis Center, California Institute of Technology, 770 South Wilson Ave., Pasadena, CA 91125, USA
            \and
            Millennium Institute of Astrophysics (MAS), Nuncio Monse{\~{n}}or S{\'{o}}tero Sanz 100, Providencia, Santiago, Chile    
             }

   \date{}

\abstract{The luminous blue variable (LBV) \rmc is located in the outskirts of the 30~Doradus complex, a region rich with interstellar material and hot luminous stars. 
We report the $3\sigma$ sub-millimetre detection of its circumstellar nebula with ALMA. The observed morphology in the sub-millimetre is different than previously observed with \emph{HST} and ATCA in the optical and centimetre wavelength regimes.
The spectral energy distribution (SED) of \rmc suggests that two emission mechanisms contribute to the sub-mm emission: optically thin bremsstrahlung and dust.
Both the extinction map and the SED are consistent with a dusty massive nebula with a dust mass of $0.055\pm0.018~M_{\odot}$ (assuming $\kappa_{850}=1.7\rm\,cm^{2}\,g^{-1}$). To date, \rmc has the most dusty LBV nebula observed in the Magellanic Clouds.  We have also re-examined the LBV classification of \rmc based on VLT/X-shooter spectra obtained in 2015/16 and a review of the publication record. The radiative transfer code CMFGEN is used to derive its fundamental stellar parameters. We find an effective temperature of $\sim 8500$~K, luminosity of log$(L/L_{\odot}) = 5.32$, and a relatively high mass-loss rate of $1.0 \times 10^{-5}~M_{\odot}$~yr$^{-1}$. The luminosity is much lower than previously thought, which implies that the current stellar mass of $\sim8~M_{\odot}$ is comparable to its nebular mass of  $\sim 5.5~M_{\odot}$ (from an assumed gas-to-dust ratio of 100), suggesting that the star has lost a large fraction of its initial mass in past LBV eruptions or binary interactions. While the star may have been hotter in the past, it is currently not hot enough to ionize its circumstellar nebula. We propose that the nebula is ionized externally by the hot stars in the 30~Doradus star-forming region. 
} 

\titlerunning{A massive nebula around the LBV \rmc revealed by ALMA}
\authorrunning{Agliozzo et al.}
\maketitle 

%

\section{Introduction}

Luminous blue variables (LBVs), also known as S\,Doradus variables, are evolved massive stars that exhibit instabilities that are not yet understood (\citealt{1984IAUS..105..233C,1997ASPC..120..387C,1994PASP..106.1025H,1997ASPC..120.....N}, and references therein). The LBV phenomenon is observed at luminosities above log$(L / L_{\odot})\sim5.2$, corresponding to stars with initial masses of $\gtrsim20~M_{\odot}$. LBVs experience outbursts with enhanced mass loss during which they appear to make transitions in the Hertzsprung-Russell (HR) diagram from their quiescent hot state ($T_{\textnormal{\scriptsize{eff}}}\sim12\,000-30\,000$~K) to lower temperatures ($T_{\textnormal{\scriptsize{eff}}}\sim8\,000$~K).  
Outbursts with visual magnitude variations of 1--2~mag are referred to as classical LBV outbursts. During giant eruptions the visual magnitude increases by more than 2~mag. Examples of stars that have experienced a giant eruption in our Galaxy are P\,Cygni in the 17th century (e.g. \citealt{1988IrAJ...18..163D,1992A&A...257..153L}) and $\eta$~Car in the 1840s (reaching $M_{V}\sim -13$, e.g. \citealt{1997ARA&A..35....1D,2012ASSL..384.....D,2011Smith}).

LBVs are considered to be stars in transition to the Wolf-Rayet (WR) stage (e.g. \citealt{1994A&A...290..819L,groh14a}, and ref. therein). 
However, recent observational and theoretical work suggests that some LBVs could be the immediate progenitors of supernovae \citep[e.g. ][]{2006Kotak,2008Trundle,2009Natur.458..865G,2007ApJ...666.1116S,2008ApJ...686..467S,2013A&A...550L...7G,groh14b,boian18}. 
Most of the fundamental questions about the physical cause of the LBV instability remain unsolved. 
Hypotheses for the mechanism involve radiation pressure instabilities, turbulent pressure instabilities, vibrations and dynamical instabilities, and binarity (see \citealt{1994PASP..106.1025H} for an overview). The high stellar luminosities near the Eddington limit probably enable instabilities (several processes in the literature are summarized in  \citealt{2002A&A...393..543V,2015ASSL..412..113O, 2014arXiv1402.0257G}). 
The impact of binarity has received much attention lately \citep{1989ASSL..157..185G,2010AIPC.1314...55K,2010ApJ...723..602K,2011MNRAS.415.2020S,2016A&A...593A..90B}.
Binary scenarios for the formation of LBVs have been proposed. LBVs may be mass gainers that received a kick when the primary exploded or the product of a merger (e.g. \citealt{2014ApJ...796..121J,2015MNRAS.447..598S,2016MNRAS.456.3401P}). However, the mass-gainer scenario presented by \citet{2015MNRAS.447..598S}, on the basis of the apparent isolation of LBVs relative to massive star clusters, has been highly debated in the literature \citep[e.g.][]{Humphreys2016}.

LBVs are surrounded by massive circumstellar nebulae of dust and gas, rich in processed material, indicative of stellar mass ejected by an evolved object through extensive stellar winds and outbursts. Representative examples in our Galaxy are the enigmatic $\eta$~Car \citep[e.g.][]{1998Morse,2003Smith,2013Smith, 2017Morris}, AG~Car \citep[e.g.][]{1950Thackeray, 2015Vamvatira}, HR~Car \citep[][]{1991Hutsemekers,2009Umana, 2017Buemi}, WRAY~15-751 \citep[][]{1991Hutsemekers,2013Vamvatira}, AFGL~2298 \citep{2001Ueta,2005Umana,2010Buemi}, and the candidate LBVs HD168625 \citep[][and ref. therein]{2010Umana}, G79.29+0.46 \citep{2008Rizzo,2011UmanaB, 2014Agliozzo, 2014Rizzo}, G26.47+0.02 \citep{2003Clark,2012Umana,2012Paron}, and the Pistol Star \citep{1999Figer,1999Lang,2014Lau}. The common presence of a N-enriched or dusty circumstellar nebula around LBVs allows the identification of candidate LBVs \citep[e.g.][]{1994PASP..106.1025H, 1995Nota, vanGenderen2001,2003Weis}, even if there is a lack of evidence of S~Doradus variability. The discovery of an infrared ring nebula around Wray 17-96, identified by \citet{2002Egan} as a candidate LBV based on the mid-IR properties of the ejecta, is  particularly important. Subsequently, the number of candidate LBVs increased considerably with mid-IR surveys carried out in the following years \citep{2005Clark, 2010Gvaramadze, 2011Wachter, 2014Flagey,2014Nowak}.

\citet{2011Kochanek} show that the physical requirement of dust condensation and growth in LBV ejecta (i.e. particle growth rate larger than grain photoevaporation rate) 
is met only for very high mass-loss rates ($>10^{-2.5}~M_{\odot}\, \rm yr^{-1}$). Such high mass-loss rates occur during LBV giant eruptions, when the stars form a cool ($\sim7000\,\rm K$) and optically thick pseudo-photosphere that shields the dust from the soft  UV photons and favours collisional particle growth with its high densities.
Dust can also form close to periastron passage in colliding wind WC binaries, like in the case of HD~193793 (see \citealt{2007ARA&A..45..177C} for a review on dust formation in WR stars). A signature of wind-wind interaction with a companion star is a dusty `pinwheel' nebula \citep{2008Tuthill}. Interestingly, a dusty spiral nebula has been observed in the binary LBV HR~Car \citep{2016A&A...593A..90B, 2017Buemi}. 

In the last few years, \emph{Spitzer} and \emph{Herschel} observations have been critical to map the dust distribution in Galactic LBV nebulae, often suggesting episodic mass-loss events or revealing the presence of a photo-dissociation region \citep[e.g.][]{2009Umana,2011UmanaA}. Mapping the dust in LBV nebulae at lower metallicities, such as in the Magellanic Clouds (MCs), has not been possible so far, because of the limited angular resolution of available infrared (IR) instrumentation. With the Atacama Large Millimeter Array (ALMA), we can for the first time map the dust of LBV nebulae in nearby galaxies in the sub-millimetre (sub-mm) bands.

\section{The LBV status of \rmc and its circumstellar nebula}
\label{sec:lbvstatus}

\subsubsection*{Is \rmc an LBV?}

\begin{figure*}
    \includegraphics[width=1\linewidth]{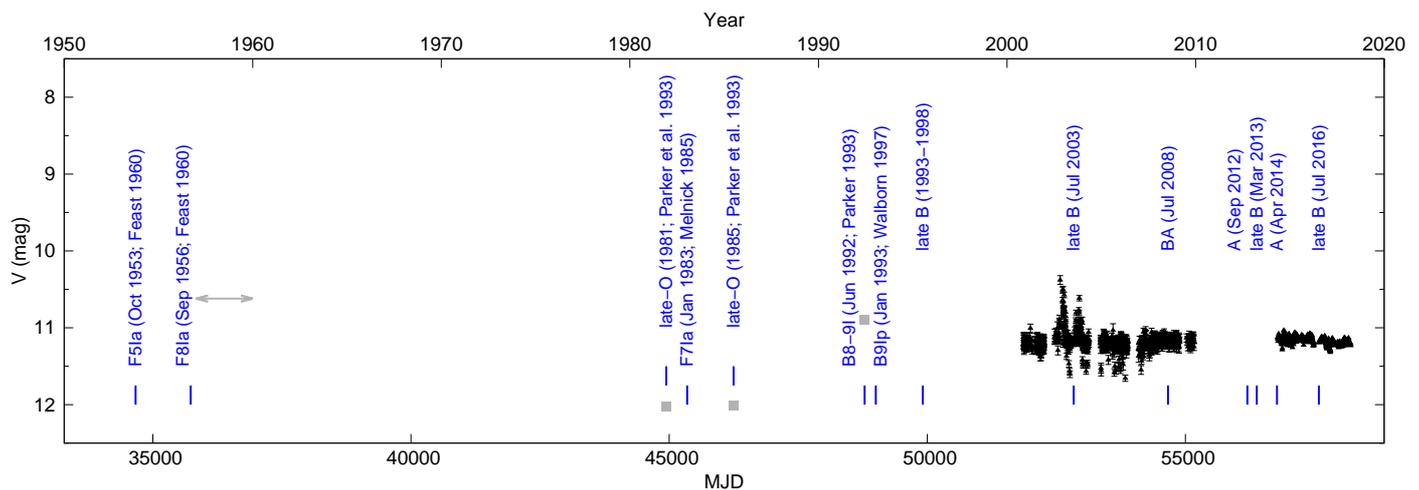}
    \caption{Lightcurve of \rmc. Historic photometry is displayed in grey, recent photometry by ASAS from 2000--2009 and ASAS-SN from 2014--2018 are displayed in black. The ASAS MAG\_4 (6 pixel aperture) is used since smaller aperture contain many outliers due to background subtraction issues. We add 0.8~mag to correct the magnitude, following \citet{2017AJ....154...15W}. We indicate the reported spectral types in blue. When no reference is given, the value has been obtained from \citet{2017AJ....154...15W}.}
    \label{fig:lightcurve}                                                                                      
\end{figure*}

\rmc\footnote{RA 05:38:51.617, Dec $-$69:08:07.315 (ICRS coord., epoch=J2000). Alternate identifier: CPD-69 463.} is classified as one of only eight confirmed LBVs in the Large Magellanic Cloud (LMC; \citealt{2018RNAAS...2c.121R}). The star lies in the outskirts of the 30~Doradus (hereafter \dor) star-forming region \citep{1997ApJS..112..457W}. 
\citet{1993Parker} classified the star as an LBV. The authors note photometric variations of at least 1.4~mag in $V$ band. Spectroscopic observations indicate that the star changed from  spectral type F5 to F8 in the 1950s, became as hot as an O9.5 star in 1981/85, before moving back to cooler temperatures and appearing as a late-B supergiant in 1992. 

The original LBV classification of \rmc by \citet{1993Parker} is questionable as the large change in $V$-band photometry reported for 1981/85 and associated change in spectral type to late-O and  also the F-type classifications in the 1950s and 1983 may be due to a misidentification of the star  (Figure \ref{fig:lightcurve}):
\begin{enumerate}
\item There has been confusion in the literature between \rmc and HD~269929\footnote{RA 05:39:37.838, Dec $-$70:39:47.042 (ICRS coord., epoch=J2000).}. In particular, \citet{1993Parker} state that HD~269929 is an alternative designation for \rmc. HD~269929 has a $V$-band magnitude of 12.2~mag and shows no significant brightness variation between 2000-2009 in the All Sky Automated Survey (ASAS; \citealt{1997AcA....47..467P}). The star is classified as F8  (SIMBAD; \citealt{1936AnHar.100..205C}), which matches the classification of \rmc in the 1950s and 1983. 
\item \citet{1985Melnick} classify stars in the centre of \dor based on spectra obtained in 1983. They assign a spectral type F7Ia, in contradiction with the O-type star classification in 1981/85 of \citet{1993Parker}. Their figure 1 identifies the correct star as \rmc.
\item \citet{1985ApJ...288..558C} exclude \rmc from their stellar sample to study interstellar dust in the LMC, as they were aware that different observers had observed two or three different stars. 
\item The 1992 June {\it International Ultraviolet Explorer (IUE)} spectrum is consistent with a B8--9I star \citep{1993Parker}. From the FES image of the {\it IUE} a V-band magnitude of 10.9~mag is estimated.
\item \citet{2017AJ....154...15W} report spectroscopic changes between the spectral types A and late-B in the time span from 1998 to 2016.
\item  Photometry variations on time scales of several weeks and of the order of 0.5~mag are observed in the ASAS lightcurve in the early 2000s. The All-Sky Automated Survey for SuperNovae (ASAS-SN) lightcurve shows no significant variations between 2014--2018 \citep{2017PASP..129j4502K,2018MNRAS.477.3145J}.
\end{enumerate}

 Unfortunately, the pre-1990 observations are not available to us. The F-type stellar classifications prior to 1990 may be observations of HD~269929 and the O-star classification in the early 1980s may also belong to observations of another star. However, the documented spectroscopic variations in \citet{2017AJ....154...15W} and the massive nebula (this work) both support an LBV classification. Given the star's relatively low luminosity for an LBV, small spectroscopic variabilities would be expected following the amplitude-luminosity relation of S~Doradus variables \citep{1989A&A...217...87W}.

\subsubsection*{The circumstellar nebula of \rmc}

The identification of a circumstellar nebula around \rmc is difficult due to its location within the \dor HII region.
\citet{1961Feast} noticed several nebular filaments, extending to an angular scale of 15\arcsec. \citet{1998Smith} analysed the velocity field and abundances of these filaments and attributed them to the \dor complex. Only a compact region up to about $2\arcsec$ from the star has abundances consistent with ejected stellar material during an LBV phase. 

High spatial resolution \textit{Hubble Space Telescope (HST)} images with the F656N filter show that the circumstellar material of \rmc is triangular shaped and is located north-west of the star \citep{2003Weis}. The nebula is elongated with an extent of 4.9\arcsec\ (about 1.2~pc) and oriented north to south-west. Its shape is very unusual and raises the speculation that strong stellar winds or turbulent motion in the HII region caused its disruption. High-dispersion optical spectra confirm the presence of a nebula with two velocity components.
 
\citet{2012Agliozzo} presented the first radio observations of \rmc, performed with the Australia Telescope Compact Array (ATCA) in Band 4cm. The resolution with the 6-km array configuration varied between 1.5\arcsec\ and 2.5\arcsec, depending on the central observing frequency (9~GHz and 5.5~GHz, respectively). This allowed to detect the radio counterpart of the optical nebula, which is ionized and emits in the radio by free-free transitions (bremsstrahlung emission). 
At 9~GHz, an unresolved source at the position of the star was detected. Comparing the radio and H$\alpha$ flux densities, which trace the same gas, led to the conclusion that the optical nebula suffers intrinsic extinction due to dust. 

In this paper, we present sub-mm/radio observations of \rmc with ALMA and ATCA and optical spectra with X-shooter at the Very Large Telescope (VLT). We perform a multiwavelength study and compare our data to archival optical {\it HST\/} Wide Field and Planetary Camera 2 (WFPC2) images and IR photometry. We interpret the nature of the ALMA sub-millimetre emission.

\section{Observations}

\begin{figure*}
\centering
    \vspace{0.2cm}
    \includegraphics[width=1\linewidth]{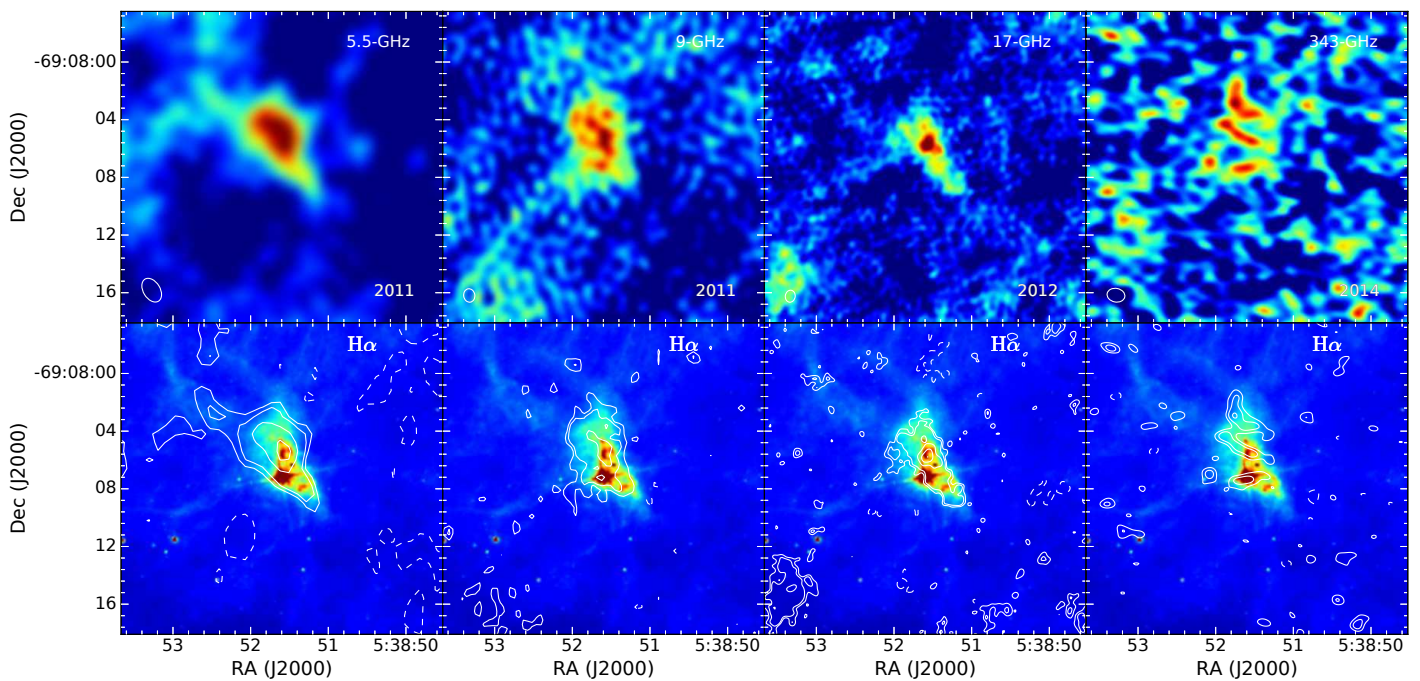}
    \caption{From top-left to top-right: ATCA 5.5, 9 and 17$\,$GHz, ALMA 343$\,$GHz maps, showing the same field of view. The synthesized beam is indicated as a white ellipse. From bottom-left to bottom-right: radio and sub-mm contours superimposed on the \textit{HST\/} H$\alpha$ image \citep{2003Weis}. The lowest contour level is 2$\sigma$, followed by 3$\sigma$ and then  increases in steps of three (steps of two at 9$\,$GHz). In the case of the 343\,GHz map, the 2, 3 and 4$\sigma$ levels are shown. All the panels are centred on the stellar position, which is the brightest source in the H$\alpha$ image.}
    \label{fig:maps}                                                                                      
\end{figure*}


%
\subsection{ALMA}
\rmc was observed with ALMA on 26 December 2014 as part of a Cycle-2 project (2013.1.00450.S, PI Agliozzo). During the observations, forty 12~m antennas were used with an integration time of 16~min. A standard Band 7 continuum spectral setup was used, resulting in four 2-GHz width spectral windows of 128 channels of XX and YY polarization correlations centred at approximately 336.5 (LSB), 338.5 (LSB), 348.5 (USB), and 350.5~ GHz (USB).  
Atmospheric conditions were marginal for the combination of frequency
and high airmass (transit elevation is $45^\circ$ for \rmc at the ALMA site). Non-standard calibration steps were required to minimize
image degradation due to phase smearing, to provide correct flux
calibration, and to maximize sensitivity by allowing inclusion of
shadowed antennas. Further discussion of these techniques and general aspects of observations and data reduction can be found in \citet{2017AgliozzoB}. We derived the intensity image from naturally weighted visibilities to maximize sensitivity and image quality (minimize the impact of phase errors on the longer baselines). We imaged all spectral windows together (343.5~GHz average; approximately 7.5~GHz usable bandwidth), yielding an rms noise of
$72\,\mu{\rm Jy}\,{\rm beam^{-1}}$.\footnote{The originally desired rms noise of $40\,\mu{\rm Jy}\,{\rm beam^{-1}}$ was not achieved because only one third of the required observations were completed.} In Figure \ref{fig:maps} we
show the 2, 3, and 4$\sigma$ contour levels of the emission. The lowest contour does not have strong statistical significance, but appears to partially overlap with the H$\alpha$ emission. Deeper observations with ALMA would improve the image quality. 

\subsection{ATCA}
\label{sec:atca}
\begin{table*}
	\centering
	\caption{Date of observation, interferometer, central frequency, largest angular scale, synthesis beam, position angle, peak flux density $F_{\nu}$, and spatially-integrated flux density $S_{\nu}$.}
	\label{tab:maps}
	\begin{tabular}{lccrcccc} 
		\hline\hline
		Date &  Array& Frequency  & LAS & HPBW & PA & $F_{\nu}$ & $S_{\nu}$ \\
		    &  & (GHz)  & (\arcsec) & (\arcsec$^2$)& (\degree) & (mJy beam$^{-1}$) & (mJy beam$^{-1}$) \\

		\hline
		2011-04-18/20& ATCA & 5.5 & 6.5\tablefootmark{a} &1.98$\times$1.60\tablefootmark{a}&$-178$&$0.51\pm0.05$&$1.3\pm0.3$\\
		2011-04-18/20& ATCA& 9 & 6.5\tablefootmark{a} &  1.30$\times$1.04\tablefootmark{a} &$-175$&$0.28\pm0.05$&$1.5\pm0.5$\\
		2012-01-21/23& ATCA& 17 & 6.5 &  0.84$\times$0.68&$-$8.8&$0.192\pm0.016$&$1.1\pm0.1$\\
		2012-01-21/23& ATCA& 23 & 4.1 &  0.63$\times$0.51&$-$8.8&<0.15&--\\
		2014-12-26 & ALMA&343 & 9.0 &  1.23$\times$0.95&\phantom{-}78.7&$0.300\pm0.072$&$2.5\pm 0.4$\\
		\hline
	\end{tabular}
\tablefoot{
\tablefootmark{(a)}  After filtering out $uv < 17\,\rm  k\lambda$ visibilities.}
\end{table*}

We performed ATCA observations of \rmc between 20 and 23 January 2012, using the array in the most extended configuration (6~km) and the Compact Array Broadband Backend (CABB) ``15~mm'' receiver in continuum mode. \rmc was observed at different hour angles distributed over three days, alternating with the phase calibrator ICRF~J052930.0$-$724528. The observations were performed in marginal weather conditions, which may have caused a loss of coherence at the long baselines (small angular scales in the sky). 
For each source we split the receiver bandwidth in two \hbox{2-GHz} sub-bands, one centred at \hbox{17\,GHz} and one at \hbox{23\,GHz}. The two data sets were reduced separately.  
Details on the observing strategy and data reduction can be found in \citet{2017AgliozzoA,2017AgliozzoB}. Information on the final images, obtained by adopting the natural weighting scheme of the visibilities, can be found in Table~\ref{tab:maps}. The table also lists information about the synthesized beam (Half Power Beam Width, HPBW), position angle (PA), largest angular scale (LAS), peak flux density, and rms-noise. 

At \hbox{17\,GHz}, we detect above $3\sigma$
the nebular emission in its whole optical extent. At \hbox{23\,GHz}, the map is noisy because of the system response to bad weather at higher frequencies and we did not detect with statistical significance emission associated with \rmc. We thus do not show the \hbox{23\,GHz} data. 

We include in our analysis the 5.5 and 9~GHz data from the ATCA observations performed in 2011 with the
CABB ``4cm-Band'' (4--10.8~GHz) receiver, described in \citet{2012Agliozzo} and shown in Figure \ref{fig:maps}. At these frequencies the HII region is bright and causes confusion. To mitigate artefacts due to the secondary lobes from bright sources in the region, 
we reprocessed the data and filtered out the $uv$ spacings smaller than $\sim 17\,\rm k\lambda$, in order to match the LAS of the 17-GHz data. This operation degrades the image quality and the rms-noise. In Table~\ref{tab:maps}, the LAS and beam information at 5.5 and 9~GHz refer to the newly processed maps, which are used to derive the spatially-integrated flux density.

\subsection{VLT/X-shooter}
\rmc was observed with X-shooter at the VLT on 2 October 2015 and 12 January 2016 (ESO programme 096.D-0047(A), PI Mehner). X-shooter is a medium-resolution echelle spectrograph that simultaneously observes the wavelength region from $3\,000$--$24\,800$~\AA\ with three arms \citep{2011A&A...536A.105V}. 
Spectra were obtained with the narrowest available slits of 0\farcs5 in the UVB arm, 0\farcs4 in the VIS arm, and 0\farcs4 in the NIR arm yielding spectral resolving powers of $R\sim9\,000-17\,000$. Spectra obtained with the 5\arcsec\ slits provide us the means to achieve absolute flux calibrations to $\sim 20$\% accuracy. However, the large aperture includes nebular reflection. The data were reduced with the ESO X-shooter pipeline version 3.2.0, and the flux normalization was computed using custom IDL routines developed by one of us (JHG). We estimate that the uncertainty in the flux normalization is about 2\%.

\subsection{{\it HST}}
The {\it HST\/} images were obtained with the WFPC2 instrument using the H$\alpha$-equivalent filter F656N (proposal ID 6540, PI Schulte-Ladbeck).
The images were retrieved from the STScI data archive, combined, and astrometrically recalibrated as described in \citet{2012Agliozzo}. These data were first published by \citet{2003Weis}.

\section{The stellar parameters and atmosphere abundances of \rmc}
\label{sec:stellarpar}

\begin{table}
\caption{Fundamental stellar parameters of \rmc from the best-fit CMFGEN model of X-shooter spectra on 2 October 2015. \label{table:stellarparameters}}
\resizebox{0.475\textwidth}{!}{
\begin{tabular}{lcc}
\hline\hline
Parameter &  Value & Error  \\ 
\hline
$T_\textnormal{\scriptsize{eff}}\tablefootmark{a} $     &   $8500$~K & $300$~K\\
$T_{*}\tablefootmark{b} $       &  9\,600~K   & 1000~K \\
log$(L_{*}/L_{\odot})$  &   $5.32$ &$0.05$ \\
$M_\textnormal{\scriptsize{bol}}$       &   $-8.56$ & $0.04$ \\
$\mathrm{BC}_\textnormal{\scriptsize{V}}$       &   $-0.14$ & $0.02$ \\
$\dot{M}\tablefootmark{c} $        & $1.0 \times 10^{-5}~M_{\odot}$~yr$^{-1}$ & $0.2 \times 10^{-5}~M_{\odot}$~yr$^{-1}$ \\
log$(g/\textnormal{[cgs]})\tablefootmark{d} $ &  $0.7$ & $0.1$ \\
$M$     & $\sim8.2~M_{\odot}$ \\
$R_\textnormal{\scriptsize{phot}}$      &  $\sim211~R_{\odot}$ \\
$R_{*}$ &  $\sim169~R_{\odot}$ \\
$v_{\infty}$    &   70~km~s$^{-1}$ &  10~km~s$^{-1}$\\
$E(B-V)$ & 0.42~mag & 0.02~mag \\
$R_V$ & 4.0 & 0.1 \\
\hline
\end{tabular}}
\tablefoot{
\tablefootmark{(a)} Parameters labelled ``$\textnormal{{eff}}$'' and ``$\textnormal{{phot}}$'' refer to $\tau_\textnormal{\scriptsize{ROSS}} = 2/3$. \\
\tablefootmark{(b)} Parameters labelled ``$\textnormal{{*}}$'' refer to $\tau_\textnormal{\scriptsize{ROSS}} = 20$. \\
\tablefootmark{(c)}  The clumping factor $f_{c}$, which is unknown, is set to unity. \\
\tablefootmark{(d)} Specified at $\tau_\textnormal{\scriptsize{ROSS}} = 2/3$. \\
}
\end{table}

\begin{table}
\caption{Surface chemical abundances for \rmc\ derived from the 2015 X-shooter spectrum.  Error estimates are on a 20\% level.  \label{table:abundances}}
\begin{tabular}{cccc}
\hline\hline 
Species & Number fraction & Mass fraction\tablefootmark{a} &  Z/Z$_{\textnormal{\scriptsize{LMC}}}$  \\ 
& (relative to H) & &   \\ 
\hline
H & 1.0 & 4.3e-01 &  0.58  \\
He & 3.3e-01 & 5.6e-01 & 2.19    \\
C & 4.0e-05 & 2.0e-04 &  0.20   \\
N & 7.5e-04 & 4.2e-03 & 14.89    \\
O & 1.5e-04 & 1.0e-03 &  0.41  \\
\hline
\end{tabular}
\tablefoot{
\tablefootmark{(a)}
LMC mass fraction used: H = 7.38e-01, He = 2.56e-01, \\
C =  9.79e-04, N = 2.82e-04, O = 2.45e-03.} 
\end{table}

Many emission lines form in the dense stellar winds of LBVs and veil the underlying photospheric spectrum. Complex radiative transfer models, which include the necessary physics to study the radiation transport across the atmosphere and wind, are needed to obtain realistic parameters.
We compare the X-shooter spectra of \rmc obtained in 2015 with atmosphere models computed with the radiative transfer code CMFGEN (version 5may17, \citealt{1998ApJ...496..407H}), which has been successfully applied to several LBVs (e.g. \citealt{2001ApJ...553..837H,2006ApJ...638L..33G,2009ApJ...698.1698G,2011ApJ...736...46G,2009A&A...507.1555C,2012A&A...541A.146C,2011AJ....142..191G,2017A&A...608A.124M}). For a review on the essential properties of the code and the basics of spectroscopic analysis of massive stars using photospheric and wind diagnostics, see \citet{2011JPhCS.328a2020G} and \citet{2011BSRSL..80...29M}. 

The diagnostics to find the best-fitting stellar model for \rmc are similar to those described in \citet{2017A&A...608A.124M}, except for the effective temperature. This quantity is constrained based on the equivalent widths of the \ion{Mg}{II} $\lambda$4481 and \ion{Mg}{I} $\lambda$5183 lines. The value of $T_{\rm eff}$ is supported by \ion{He}{I} $\lambda$4713, \ion{He}{I} $\lambda$5876, photospheric \ion{Fe}{II} lines, and the emission components of \ion{Fe}{II} lines formed in the wind. The mass-loss rate is constrained based on the emission components of \ion{Fe}{II} and H Balmer lines. We note that in this parameter regime, the strength of the H emission lines is extremely sensitive on the effective temperature and temperature structure of the wind. 
The luminosity of \rmc is constrained based on the absolute flux calibration of the X-shooter spectrum used to constrain the other stellar parameters. Since \rmc is a variable star, we consider this more reliable than using broad-band photometry obtained at different epochs.

\begin{figure}
\centering
\includegraphics[width =1.0\columnwidth]{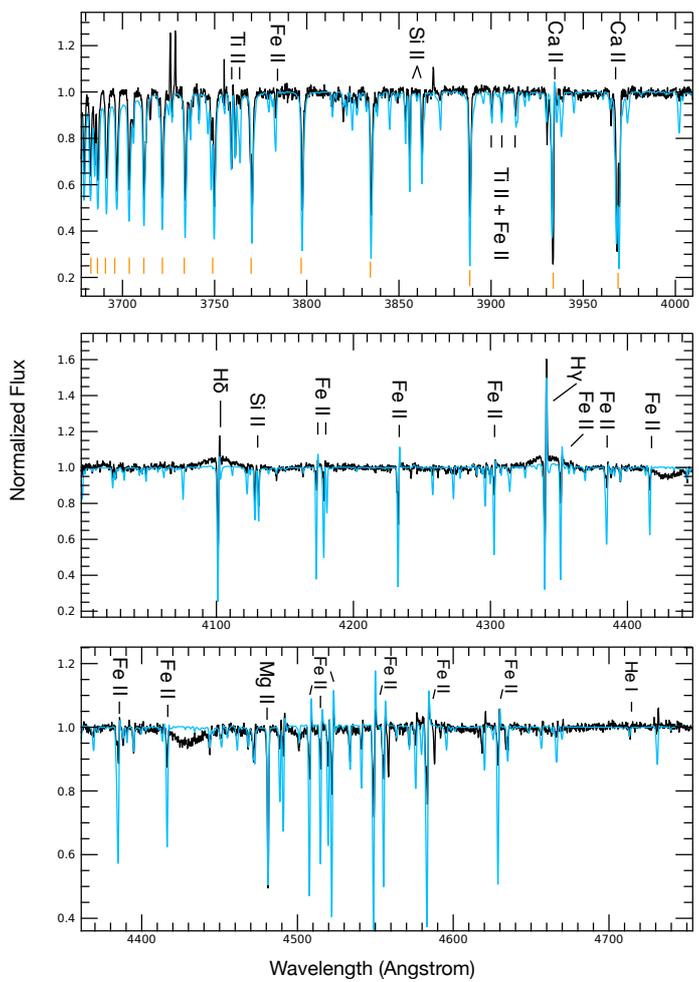}
\caption{2015 X-shooter spectrum of \rmc (black line) and best-fit CMFGEN model (blue line). The spectral ranges cover important hydrogen and heliums lines, but also \ion{Si}{II}, \ion{Mg}{II}, \ion{Fe}{II}, and \ion{O}{I}. The orange ticks correspond to hydrogen lines.}
\label{fig:cmfgenspec1}
\end{figure}

\begin{figure}
\centering
\includegraphics[width =1.0\columnwidth]{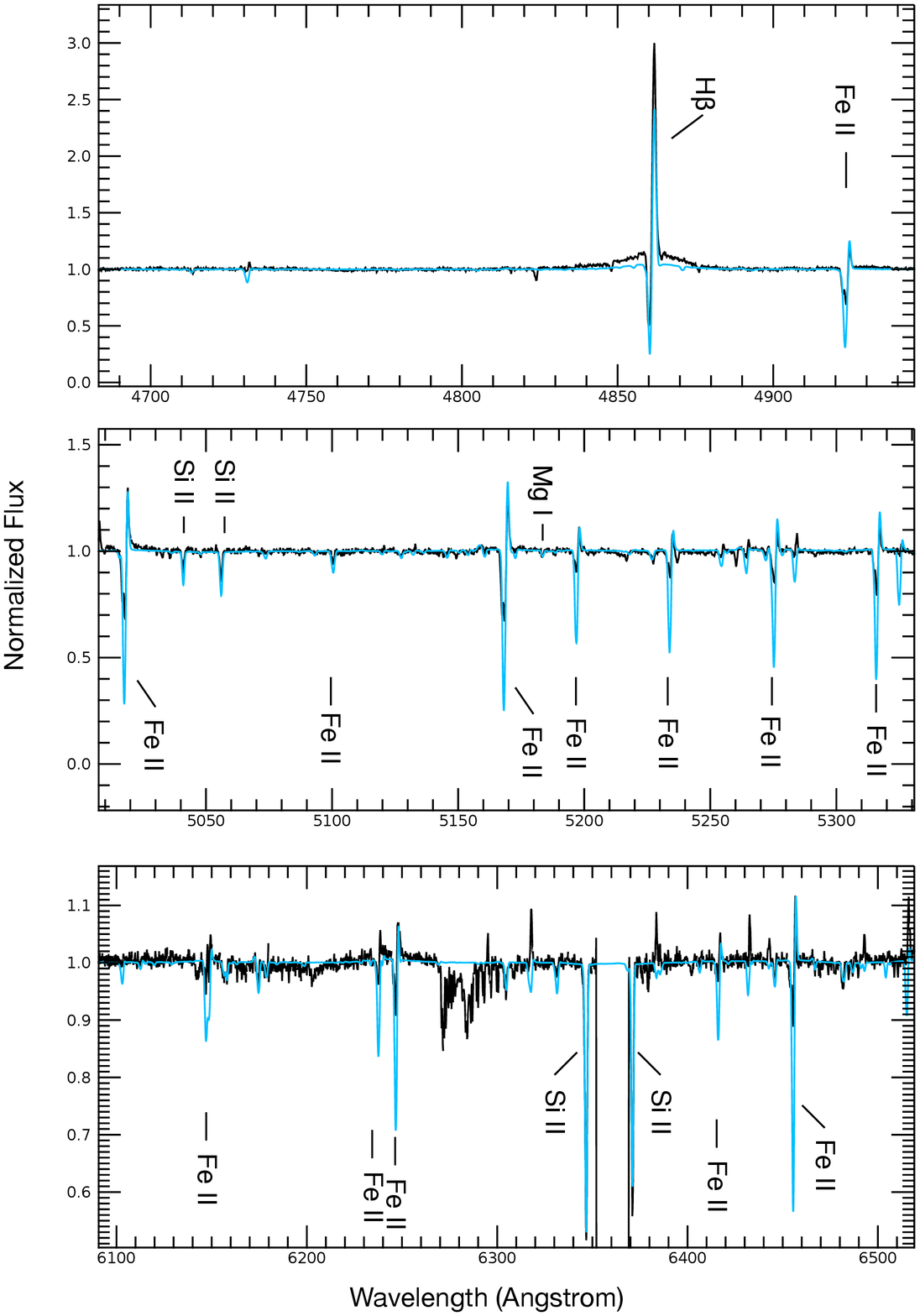}
\caption{Like Figure \ref{fig:cmfgenspec1}.}
\label{fig:cmfgenspec2}
\end{figure}

\begin{figure}
\centering
\includegraphics[width =1.0\columnwidth]{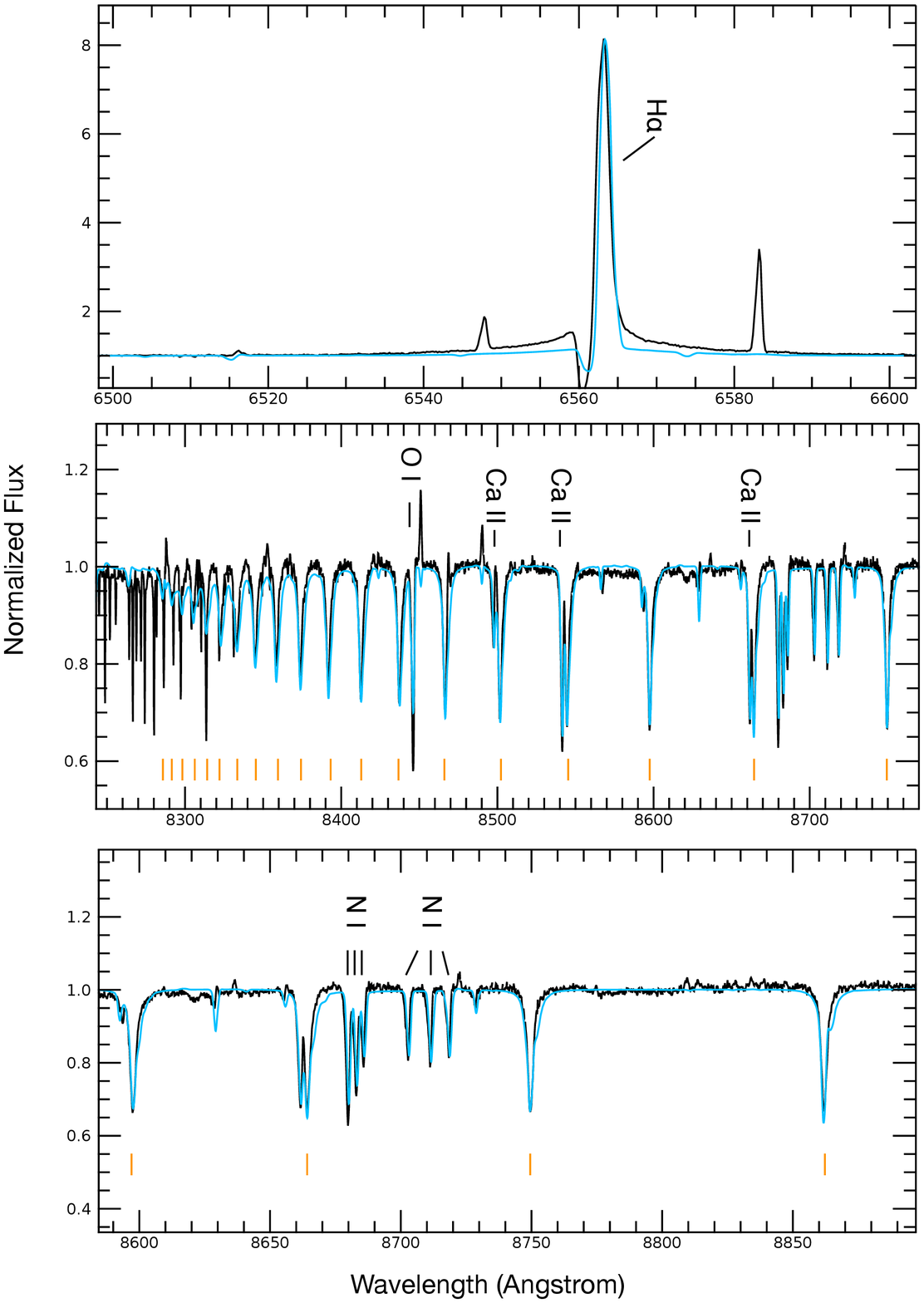}

\caption{Like Figure \ref{fig:cmfgenspec1}.}
\label{fig:cmfgenspec3}
\end{figure}

Figures \ref{fig:cmfgenspec1}, \ref{fig:cmfgenspec2}, and \ref{fig:cmfgenspec3} compare the 2015 X-shooter spectrum of \rmc with our best-fit CMFGEN model. The derived stellar parameters for its quiescent state are listed in Table \ref{table:stellarparameters}.
We find an effective temperature of $T_\textnormal{\scriptsize{eff}} = 8\,500 \pm 300$~K and a stellar luminosity of log$(L/L_{\odot}) = 5.32 \pm 0.05$ ($M_\textnormal{\scriptsize{bol}}  = -8.56$~mag).
We note that the value of $T_\textnormal{\scriptsize{eff}}$ derived from the \ion{Mg}{I}/\ion{Mg}{II} line ratio does not provide enough free electrons in the wind, which are needed for producing the electron scattering wings clearly visible in the H recombination lines. In addition, the H ionization structure seems to require a slightly higher value of $T_\textnormal{\scriptsize{eff}}$. In this parameter range, our models show that a small increase of $T_\textnormal{\scriptsize{eff}}$ (by 500-800 K) would be enough for producing stronger H emission and broader electron scattering wings. We still prefer to give higher weight to the \ion{Mg}{I}/\ion{Mg}{II}, since these lines are formed deeper in the wind. Our interpretation is that the H ionization structure is not fully reproduced by our models, and is likely affected by time-dependent effects that are inherent to LBVs \citep{GrohVink2011}.

\begin{figure}
\centering
\includegraphics[width =1.0\columnwidth]{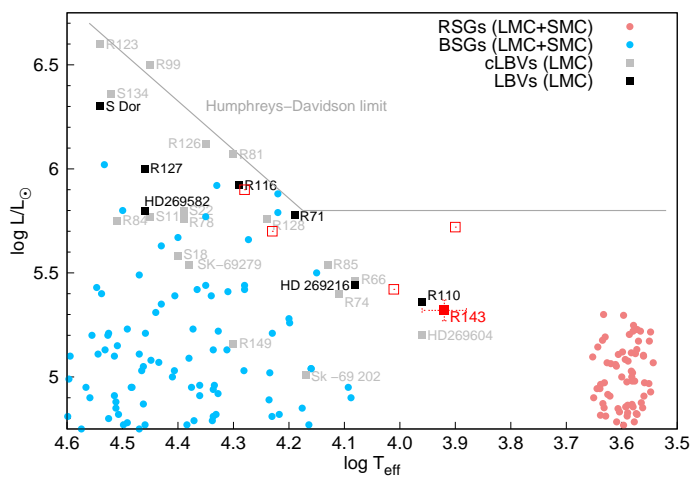}
\caption{Schematic upper HR diagram of the LMC, adapted from \citet{2017A&A...608A.124M}. Confirmed LBVs (black) and candidate LBVs (grey) are shown at their minimum phases. Blue and red supergiants are indicated with blue and red filled circles. We show the location of \rmc obtained in this work (filled red square). Previous literature values for \rmc are also shown (open red squares).}
\label{fig:HRdiagram}
\end{figure}

Our derived value of log$(L/L_{\odot}) = 5.32$ is lower than the most commonly-used estimate (log$(L/L_{\odot}) = 5.7$), which was suggested by \citet{1993Parker}. However, our luminosity value is in line with the estimate from \citet{1997ApJS..112..457W}, who originally suggested that R143 had an initial single-star mass around $25-30~M_\odot$. The two most likely possibilities are that (1) \citet{1993Parker} misidentified \rmc (see Section \ref{sec:lbvstatus}), or (2) \rmc had a different value of luminosity and temperature around the early 1990s. Figure \ref{fig:HRdiagram} shows the position of \rmc in the HR diagram obtained from our CMFGEN modelling.

Extended emission from surrounding material is apparent in the spectra and not all hydrogen and helium lines are matched equally well. The estimation of the effective temperature $T_{\rm eff}$ is not affected by the nebular contamination, while the luminosity $L$ may be somewhat overestimated due to inclusion of reflected or scattered nebular continuum. Since the hydrogen absorption lines are filled in with extra emission from the surrounding nebula, we only derive a lower limit of the terminal wind velocity $v_{\infty} > 70\,\rm km\,s^{-1}$.

The gravity is estimated to be log$(g/\textnormal{[cgs]}) = 0.7$ and the stellar mass to be $\sim8~M_{\odot}$. A mass-loss rate of $\dot{M} \sim 1.0 \times 10^{-5}~M_{\odot}$~yr$^{-1}$ for a clumping factor $f_{c}$ equal to 1 fits the hydrogen emission well. The mass-loss rate is more affected by the uncertain ionization structure and time dependence than by nebular contamination since we used mainly H$_\beta$, higher Balmer lines, and \ion{Fe}{II} lines to constrain it.  For comparison, the LBV W243 with a similar temperature of $T_{\mathrm{eff}} \sim 8500$~K but much higher luminosity has a much weaker mass-loss rate of $6.1 \times 10^{-7}~M_{\odot}~$yr$^{-1}$ \citep{2009A&A...507.1597R}.

\rmc displays the spectrum of an LBV. In particular, the spectrum displays no [O I] $\lambda\lambda$6300,6364 emission, characteristic of the sgB[e] class \citep{2017ApJ...836...64H}. We also detect no CO first-overtone emission at $2.3~\mu \rm m$, indicative that \rmc has no high density circumstellar disc as is the case for B[e] stars \citep{1988ApJ...334..639M}. 
\rmc shows signs of CNO-processed material at the surface (Table \ref{table:abundances}), with enhanced He and N, and depleted H, C, and O. This confirms the evolved nature of the star, especially given the low current mass ($\sim 8~M_{\odot}$).

\section{Morphology of the radio and sub-millimetre emission}
 \label{sec:morphology}
\begin{figure*}
        \centering
        \includegraphics[width=0.9\textwidth]{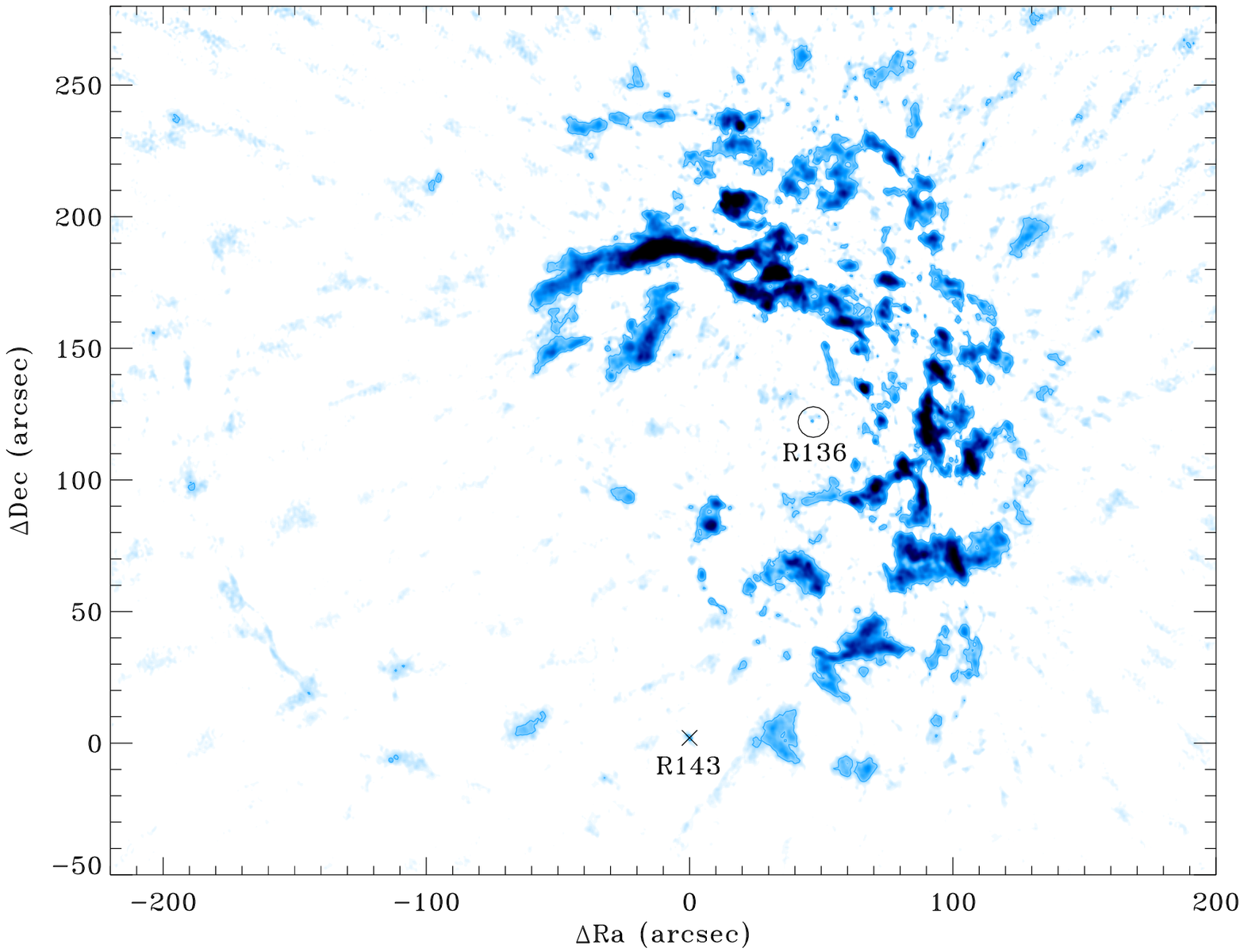}
    \caption{ATCA 5~GHz map from the observations presented in \citet{2012Agliozzo}, showing the 3-$\sigma$ contours. The cross indicates the position of \rmc, which is the phase centre of the field, and the circle the position of the stellar cluster RMC$\,$136. The primary beam FWHM is about 9~arcmin.}
\label{fig:r143_5GHz}
\end{figure*}

\begin{figure}
    \begin{minipage}{1\linewidth}
        \centering
        \includegraphics[width=1\textwidth]{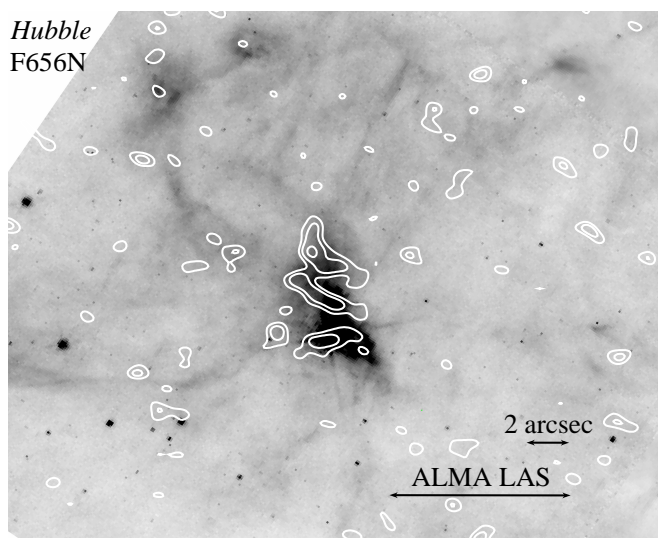}
        \caption{\emph{HST}/H$\alpha$ image \citep{2003Weis}, overlaid with the $2, 3, 4\sigma$ contours of 
      the ALMA map at \hbox{343\,GHz}. North is up and east is left.}
        \label{fig:hst-alma}
    \end{minipage}
\end{figure}
 
\begin{figure*}
\centering
        \includegraphics[width=\textwidth]{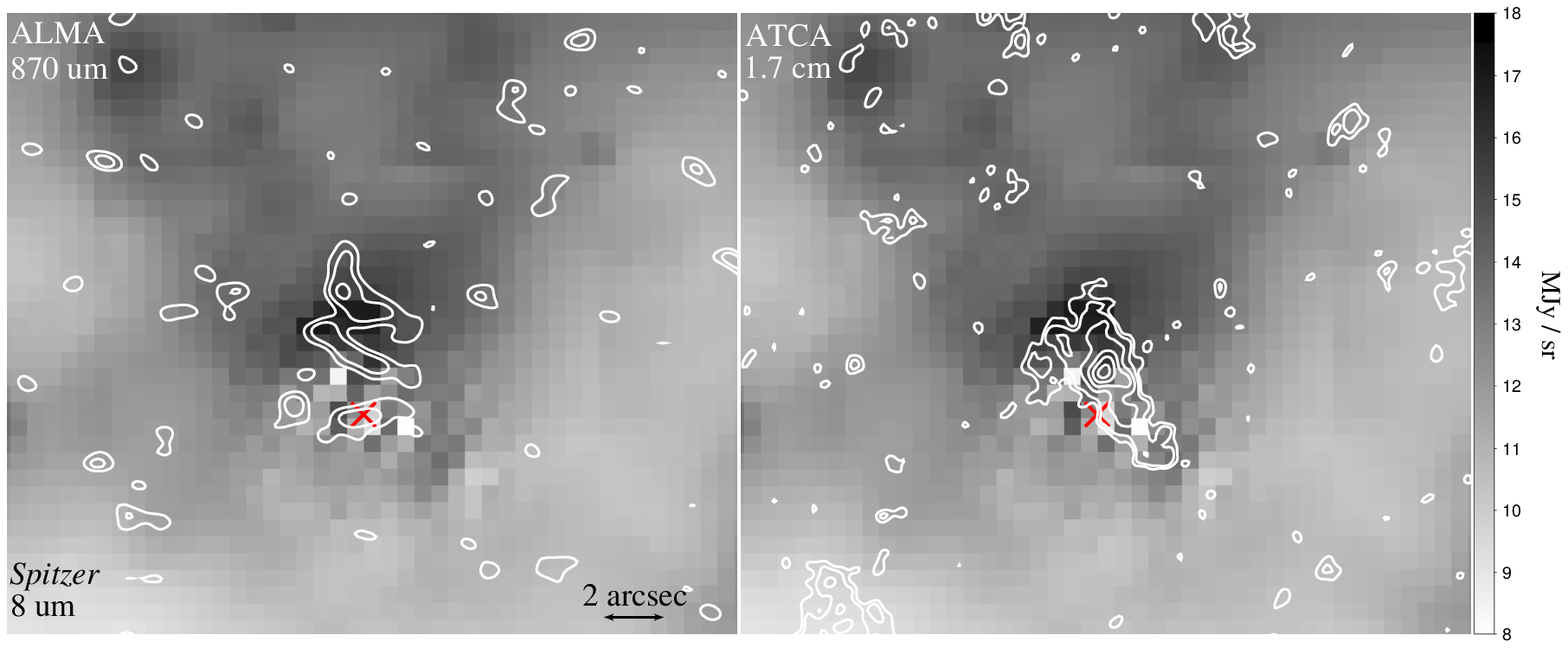}
        \caption{Residuals from the 8 $\rm \mu m$ IRAC image after point-source subtraction. The white contours are the [2, 3, 4 $\sigma$] ALMA Band 7 emission (left) and the [2, 3, 4, 5, 6, 7 $\sigma$] ATCA 17 GHz emission (right). North is up and east is left. The red cross indicates the  position of the star.}
        \label{fig:IRACresiduals}
\end{figure*}
 
\rmc lies south-east from the centre of the \dor star-forming region, the most luminous HII region in the Local Group. The environment around \rmc is rich in interstellar material and hot luminous stars and is very bright in the optical, IR, and radio.  Figure \ref{fig:r143_5GHz} shows part of the \dor complex as seen by ATCA in the 6-km configuration at 5~GHz. The LAS in this map is  $\sim20\arcsec$, which means that more extended radio emission in the star-forming region is filtered out by the interferometer. The Feast filaments (with an angular scale of $\sim15\arcsec$) are not detected at 5~GHz, probably because they are too faint, although the $2\rm \sigma$ contour elongated toward the N-E side of the nebula suggests a partial detection of one of the filaments. At higher frequencies, the filaments would be filtered out because of the lack of baselines for those angular scales. The Feast filaments and \rmc's nebula are illustrated in the H$\alpha$ image of Figure \ref{fig:hst-alma}, with the ALMA contours at 343~GHz overlaid. 

As is the case in the optical, the nebula is also triangular-shaped in the radio, with no counterpart in the east and south-east of the star (Figure \ref{fig:maps}). From the \hbox{17\,GHz} map, which has the highest spatial resolution, we estimate a size (beam convolved)  of the radio emission of about 6\arcsec\ along the largest extension.  
The point-source detected at \hbox{9\,GHz} at the position of the star is not detected at \hbox{17\,GHz}, most likely because of lack of coherence at the smallest angular scales due to high phase-rms during the observations. The knot of bright emission visible at all radio frequencies corresponds to the nebular material 2\arcsec\ north of the star, attributed to \rmc \citep{1998Smith}. 

The sub-mm emission detected by ALMA has an irregular shape and is distributed on one side of the star (Figure \ref{fig:hst-alma}), similarly to the optical and radio emissions. However, the sub-mm contours suggest a different morphology and thus a different nature of the emission. The sub-mm continuum emission can be due to both free-free emission from the ionized gas and to thermal emission from dust. The slightly curved emission in the sub-mm seems aligned with an ``imaginary'' extension of at least one of the Feast filaments over \rmc's nebula. This is not seen in the radio maps. 
A compact object at the position of the star is also detected in the ALMA map with a statistical significance of $4\sigma$. 

Ideally, we would like to compare the sub-mm continuum emission with far-IR images. Unfortunately, the spatial resolution of IR space telescopes is unsuitable for disentangling the contribution of the central star, the circumstellar nebula, and the interstellar dust from the total emission. For example, the {\it Spitzer\/}/MIPS 24 $\rm \mu m$ image has a resolution of $\sim 6\arcsec$, too poor for morphological comparison with other wavelengths.
We retrieved the 8 $\rm \mu m$ IRAC image \citep[][with a resolution of $\sim 2\arcsec$]{2006Meixner} and subtracted the point source at the position of \rmc using the tool APEX in the MOPEX package \citep{2005Mopex,2011Mopex}. The point-source subtraction worked relatively well considering the high and not uniform background, and the residual image is sufficient for a qualitative comparison with the ALMA and ATCA data.
The spatial resolution is still poor and the central part of the nebula may have been subtracted together with the point-source. However, it is possible to recognise an outer asymmetric emission that seems  to overlap better with the sub-mm 343~GHz ALMA contours than the cm 17~GHz ATCA contours (Figure \ref{fig:IRACresiduals}).
Thus, the $\rm 8~\mu m$ flux plausibly arises from the same component responsible for the ALMA contours.

\begin{figure*}
    \centering
    \includegraphics[width=1\linewidth]{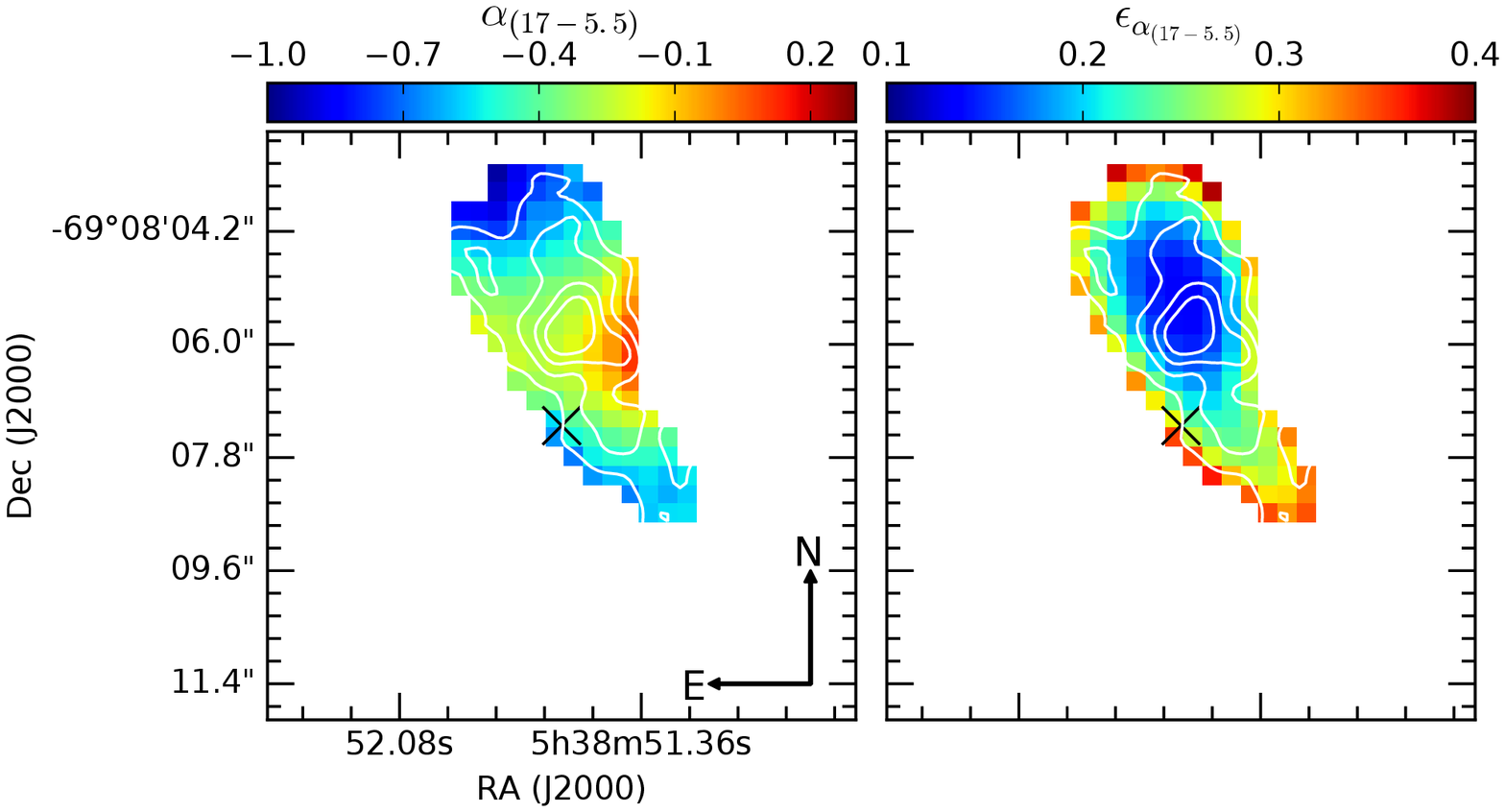}
    \caption{Left: Spectral index map between 17 and \hbox{5.5\,GHz}. Right: Error spectral index map in Jy~pixel$^{-1}$. The map
    at 17~GHz was reconvolved in order to match the beam at 5.5~GHz. The white contours indicate flux densities above 3, 5, 7, 9 $\sigma$ of the 17~GHz emission. The cross indicates the position of the star. }
    \label{fig:spix_lbvs}
\end{figure*}

\section{Ionized matter}
\label{sec:ionized}

The flux density of \rmc and the rms-noise $\sigma$ is determined from the \hbox{17\,GHz} radio map using the CASA \texttt{viewer}. For the flux density, we integrate the nebular emission of areas above $3\sigma$ level. The rms-noise is evaluated in regions free of emission and the flux density error is estimated as $\epsilon=\sigma \sqrt{N}$, where $N$ is the number of independent beams in the selected region. Calibration errors are negligible and the uncertainty is dominated by the noise in the maps and possible systematic under-estimated due to the $3\sigma$ cut. 
We also estimate the flux densities at 5.5 and \hbox{9\,GHz} from the reprocessed maps (Section \ref{sec:atca}) to reduce the contribution from the HII region extended emission and/or image artefacts due to zero-baseline scale missing \citep[Table \ref{tab:maps}; compare with][]{2012Agliozzo}. In fact, both can contaminate the nebular emission and eventually falsify the flux density analysis. 
Because of the inhomogeneous background in the 5.5 and 9~GHz maps, the flux density measurements have a larger uncertainty, especially at 9~GHz. 

The total integrated flux density encompassing the nebula and the central object at 343~GHz is $2.5\pm0.4\,\rm mJy$ (the error also includes 10\% flux uncertainty). Part of this emission probably originates from the ionized nebula and is due to bremsstrahlung. In fact, from the flux density at 17~GHz, the extrapolated value at 343~GHz accounting for the free-free emission is $\sim0.8\,\rm mJy$, about one third of the observed total flux density at 343~GHz.

We derived a spectral index map (per-pixel) between 5.5 and \hbox{17\,GHz}, after re-gridding the highest-resolution map (\hbox{17\,GHz}) to the same grid of the \hbox{5.5\,GHz} map and re-convolved it with a synthesized beam that matches the resolution at \hbox{5.5\,GHz}, similar to \citet{2017AgliozzoA}. Figure \ref{fig:spix_lbvs} shows the spectral index map and its associated error map. Calibration errors are negligible and the error in each pixel is given by the sum in quadrature of the rms-noise in both the maps. The weighted fit of the power-law $S_{\nu}\propto\nu^{\alpha}$ between the flux densities at 5.5, 9, and \hbox{17\,GHz} gives us a mean spectral index $\langle\alpha\rangle=-0.2\pm0.2$, which is consistent with optically thin free-free emission typical of evolved HII regions (theoretically it is $S_{\nu}\propto\nu^{-0.1}$). 
The knot visible at all the radio frequencies in the nearly-middle part of the nebula has a flat flux density distribution (spectral index $\alpha\approx 0$), typical of optically thin bremsstrahlung emission. This suggests that its large brightness relative to the surrounding nebula is due to additional material along the line of sight rather than a density clump.

The point-source at the position of the star visible at 9 and 343~GHz has peak flux densities of $0.20\pm0.05\,\rm mJy\,\rm beam^{-1}$ and $0.27\pm0.07\,\rm mJy\,\rm beam^{-1}$, respectively.  
We use the stellar parameters derived in Section \ref{sec:stellarpar} and the formulation of the spectrum of free-free radiation from ionized spherical stellar winds by \citet{1975PF} to determine the expected flux density at 9, 17 and 343 GHz, as
\begin{align}
&S_{\nu} = 3\times10^{10}\, \nu^{0.6}_{[\rm GHz]}\, T_{e\,[\rm K]}^{0.1}\, (\dot{M}_{[M_{\odot}\,\rm yr^{-1}]}\sqrt{f_{c}})^{4/3}  (v_{\infty \,\rm [km\,s^{-1}]}\times\, \mu)^{-4/3}  \nonumber\\
&\quad \quad \times        \,\overline{Z}^{2/3}\, D^{-2}_{\rm [kpc]}\, [\rm mJy],  \nonumber
\end{align}
where we assume a mean molecular weight per ion $\mu$ equal to 1.33 (adopting the He abundance determined in Section \ref{sec:stellarpar}), an electron temperature of $T_{e} = 5\,000\,\rm K$ and an average ionic charge of $\overline{Z}=0.9$, accounting for the fact that in such a cool star He should be neutral. Same values for $T_{e}$ and $\overline{Z}$ were used by \citet{1995Leitherer} to model the radio emission of Cygnus OB2 No.\ 12. We set $f_{c}$ equal to 1. With these assumptions we find flux densities of $0.45^{+0.13}_{-0.10}\,\rm mJy$, $0.07\pm0.02\,\rm mJy$ and $0.05^{+0.02}_{-0.01}\,\rm mJy$ at 343, 17 and 9 GHz, respectively. The errors were derived taking into account the uncertainties from the CMFGEN model and a $10\%$ uncertainty for $\overline{Z}$ and $\mu$. Based on this model, at 343 GHz we would expect to observe a higher flux density; at 9 GHz the measurement might include some nebular flux; at 17 GHz we might have lost the detection of the point source because of decorrelation during the observations due to unstable weather. However, the terminal velocity $v_{\infty}$ determined in Section \ref{sec:stellarpar} is a lower limit, thus the above flux densities must be treated as upper limits. 

It seems plausible that the sub-mm emission is due to free-free, but such a model of stationary, isothermal and spherically symmetric optically thick ionized wind may not applicable to \rmc. The ionization fraction in the wind might be lower than assumed, given the low effective temperature and the hydrogen in the wind might recombine at large distances from the surface, among many possible scenarios. For $\rm RMC\,127$, \citet{2017AgliozzoB} found that a collimated ionized wind model better explains the observed SED from the near-IR to the radio.
 Other measurements in between the two ALMA and ATCA frequencies and between the near-IR and the sub-mm would be very useful to better understand the nature of the central object emission. Another source of emission in the sub-mm could be dust in the wind or in a shell in front of the star. 

\section{The spatial distribution of dust}

\begin{figure}
    \includegraphics[width=0.48\textwidth]{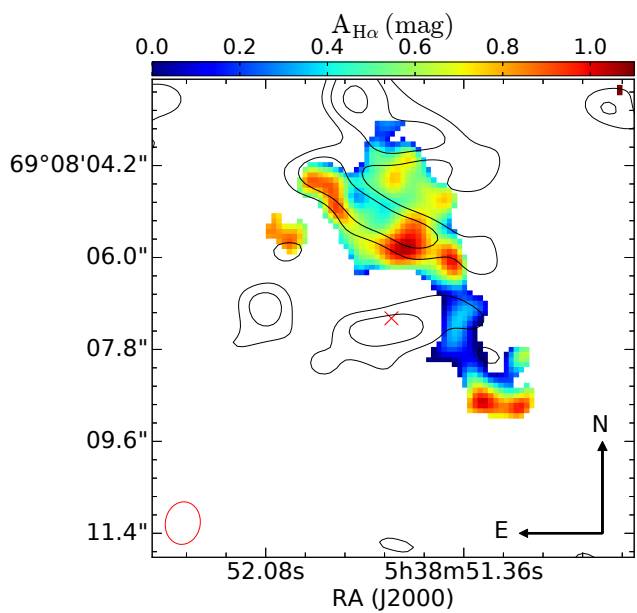}
    \caption{Extinction map in H$\alpha$ derived by comparing the  H$\alpha$ recombination line and the centimetre (\hbox{17\,GHz}) emission above $3\sigma$. The central star, indicated with the red cross, is masked with a circular aperture in the optical image. 
The sub-mm contours at \hbox{343\,GHz} are shown (black lines). } 
    \label{fig:ext1}
\end{figure}

The H$\alpha$ emission in the nebula of \rmc is due to the de-excitation of the recombined hydrogen atom and the radio centimetric emission is due to free-free transitions. They trace the same gas \citep{2012Agliozzo} and it is thus possible to determine the extinction of the H$\alpha$ line by comparing the surface brightness at optical and radio wavelengths \citep{1984Pottasch}. 

We follow the procedure described in \citet{2017AgliozzoA}. An electron temperature of $12\,200\,\rm K$ \citep{1998Smith} is adopted. The extinction map is derived by comparing pixel-by-pixel the highest-resolution radio image (\hbox{17\,GHz}) with the {\it HST\/} $\rm H \rm \alpha$ image as
$2.5\,\log(F_{\rm 17\,GHz\,(obs)}/F_{\rm 17\,GHz\,(expected)})$, where $F_{\rm 17\,GHz\,(expected)}$ is the expected radio emission from the $\rm H \rm \alpha$ recombination-line 
emission \citep[see equations 7 and 8 in][]{2017AgliozzoA}. This computation was performed in
every common pixel with brightness above 3$\sigma$, where $\sigma$ was computed by summing in quadrature the noise in the maps and
calibration uncertainties. We mask the
$ \rm H \rm \alpha$ emission from the star, keeping in mind that we want to estimate the expected
free-free emission from the optical nebular line. The derived extinction map is illustrated in
Figure \ref{fig:ext1}. 

Relatively small extinction due to dust is evident across the entire region. The maximum value for $A_{\rm H\alpha}$ is $\sim$1.06~mag and corresponds to the knot of the radio and optical emission. This knot is not a density clump according to the spectral index analysis and is likely the effect of material along the line of sight.  The most extinguished region is in the same direction of the ALMA emission corresponding to the middle part of the nebula. The ALMA emission may arise mostly from a cool dust component that surrounds the ionized gas in the nebula and extinguishes the optical $\rm H\alpha$ emission.

\section{The SED of \rmc}
\begin{figure*}
    \centerline{\includegraphics[width=0.9\textwidth]{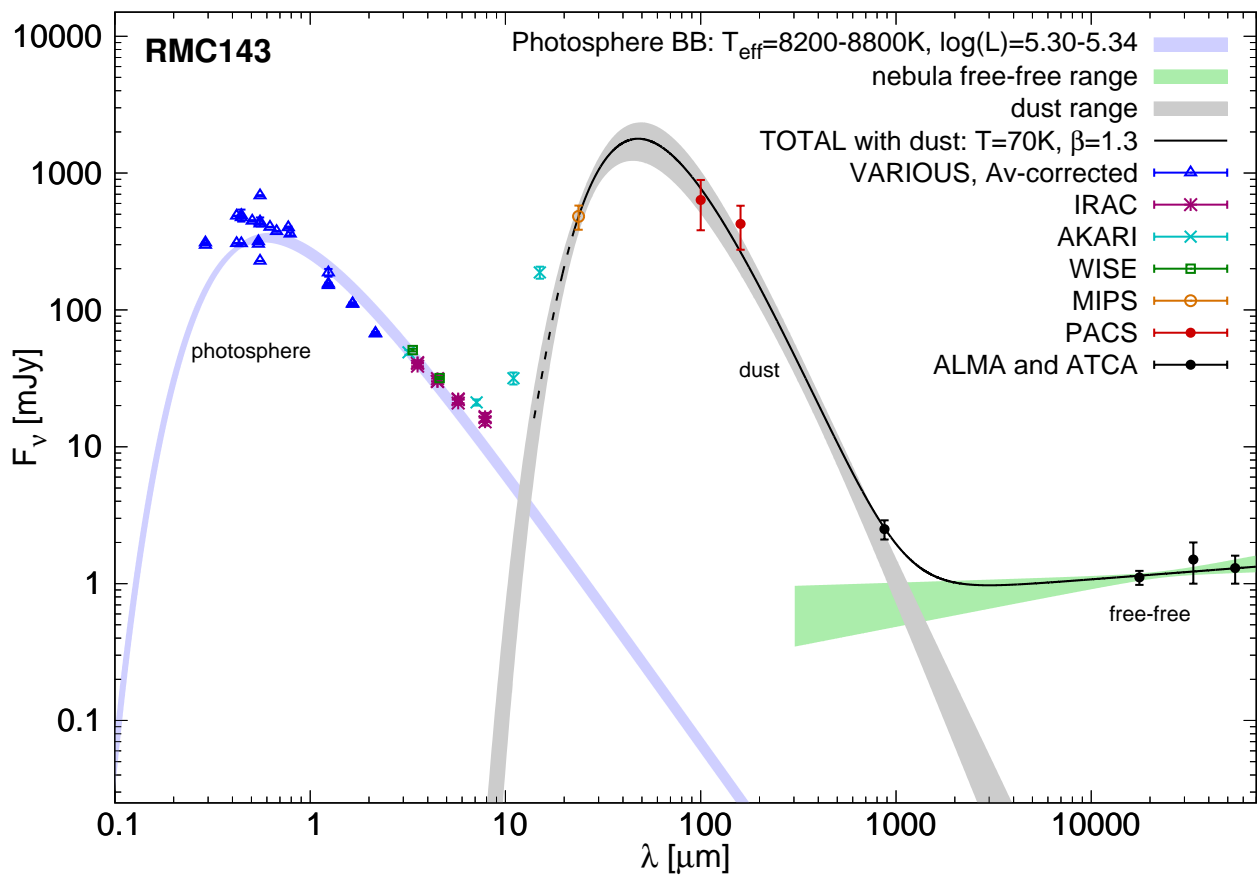}}
    \caption{Flux density distribution of \rmc from IR catalogues and our ALMA and ATCA measurements. The dust component shown is a greybody fit to the MIPS24, PACS100, PACS160, and ALMA photometry. The grey and green bands show the ranges of values within twice the minimum $\chi^2$ for the dust greybody and nebula free-free fits, respectively. We note that the AKARI S11 and L15 photometry have a colour correction applied assuming a $T=70\,\rm K$, $\beta=1$ greybody.}
    \label{fig:sed}
\end{figure*}

\begin{table}
\centering
\caption{Photometry used in the dust greybody fit.}
\label{tab:photometry}
\begin{tabular}{llccrcl}
   \hline\hline\\
   Telescope&Instr.&$\lambda$&FWHM&\multicolumn{3}{c}{$S_\nu$} \\
   & & [$\mu \rm m$]& [$''$]&\multicolumn{3}{c}{[mJy]}\\
   \hline
   \emph{Spitzer} &MIPS  &23.675& 5.9&482.5&$\!\!\!\!\pm\!\!\!\!$&96.5\tablefootmark{a} \\
   \emph{Herschel}&PACS  &100   & 6.7&636  &$\!\!\!\!\pm\!\!\!\!$&254\\  
   \emph{Herschel}&PACS  &160   & 11 &426  &$\!\!\!\!\pm\!\!\!\!$&151\\
   ALMA           &Band 7&872.8 &1.1\tablefootmark{b} & 2.5 &$\!\!\!\!\pm\!\!\!\!$&0.4\\
    \hline
\end{tabular}
\tablefoot{
\tablefootmark{(a)} MIPS24 uncertainty set to 20\% to account for confusion. \\
\tablefootmark{(b)} ALMA measurement is integrated over the nebula.
}
\end{table}

Our ALMA sub-mm observations can  constrain some of the dust properties. To understand the origin of the ALMA sub-mm emission we compare it with the available IR photometry, which were derived from images with different spatial resolutions.  We consulted the IR catalogues with the CDS VizieR \citep{2000Ochsenbein} and IRSA Gator\footnote{\tt https://irsa.ipac.caltech.edu/applications/Gator/} tools. For the mid- and far-IR photometry we evaluated the effect of confusion on photometry. We include photometry from the 2MASS Point Source Catalog \citep{2003Cutri}, the \textsl{Spitzer} SAGE legacy survey \citep[Data Release 3/final]{2006Meixner}, the \textsl{AKARI} IRC LMC survey \citep[][]{2012Kato,2008Ita}, \textsl{WISE} \citep{2012Cutri}, and \textsl{Herschel} PACS \citep{2013Meixner}. We include our measurements of the ALMA Band 7 and ATCA Bands 4cm (5.5 and 9 GHz) and 15mm (17GHz).

We have excluded the \textsl{WISE} W3 and W4 band photometry due to confusion in the relatively large effective beams ($\sim\!7''$ for W3 and $\sim\!12''$ for W4, $\sqrt{2}$ larger in Atlas images) and the complication of the very wide W3 spectral response. The W4 band photometry in particular is highly discrepant with the MIPS24 photometry at similar wavelength but much higher spatial resolution ($5.9''$ FWHM). We consider as valid the AKARI S11 and L15 band photometry. Although the AKARI all-sky survey does not have sufficient spatial resolution, the LMC survey was taken in a pointed imaging mode which yields comparable resolution to MIPS24 in the L15 band and even better in the S11 band \citep[FWHM of $5.7''$ for L15 and $4.8''$ for S11,][]{2007Onaka}. The available \textsl{Herschel} PACS100 and 160 band photometry is somewhat affected by confusion due to the large PSFs ($7.7''$ and $12''$ FWHM), but this appears to be well accounted for in the large quoted error bars, and they are the only far-IR photometry available. Unfortunately, there is no PACS70 data for \rmc, which would have had a better spatial resolution and been closer to the peak of the dust emission. There is no detection by MIPS70 due to the very large beam ($18''$ FWHM) and confusion. For MIPS24 we take the weighted average of the two SAGE epoch measurements of $2.974\pm0.016\,\rm{mag}$ and $2.913\pm0.010\,\rm{mag}$, that is,  $2.930\,\rm{mag}$ or $482.5\,\rm{mJy}$. Due to the expected confusion we assume a 20\% uncertainty on the MIPS24 photometry.

We also include the photometry available in VizieR from the UV and optical catalogues XMM-Newton, AC2000, AAVSO, GAIA, and RAVE  \citep{2012Page,1998Urban,2015Henden,2016Gaia, 2017Kunder}. We correct these data for interstellar dust extinction, using the extinction curve from \citet{1999PASP..111...63F},  $R_\mathrm{V}=4.0$ and $E(B-V) = 0.42$~mag, as determined from our CMFGEN modelling (see Table \ref{table:stellarparameters}). Keeping in mind that the UV to near-IR photometry comes from observations performed at different epochs and the star being likely variable, we can argue that the corrected data (blue triangles in Figure \ref{fig:sed}) agree with the CMFGEN model of a star with  $T_\mathrm{eff}=8\,500\pm300\,\rm K$, log$(L_{*}/L_{\odot})=5.32$ at a distance of $49.97\,\rm kpc$ (Table \ref{table:stellarparameters}).

The SED of \rmc's circumstellar material can be described with two components: 1) the ionized gas in the nebula that can be modelled by optically thin bremsstrahlung dominating the total flux at radio wavelengths (see Section \ref{sec:ionized}); 2) a dust component that peaks in the far-IR, typical for LBV nebulae \citep[e.g.][]{1997Hutsemekers,2003Clark}.

Given the limitations of the available photometry and lack of a mid-IR spectrum, we fit only a simple single-temperature greybody \citep[power-law opacity index $\beta$, see e.g.][]{1993backman} to model the dust and obtain an estimate of the dust mass. As wavelengths shortward of about $20~\mu\rm m$ can deviate strongly from a simple greybody due to PAH and/or silicate spectral features  \citep[typically in HII regions, but also in the nebula of the candidate LBV HD\,168625,][]{2007Draine,2010Compiegne,2010Umana} or due to several ionized gas forbidden lines \citep[e.g. as seen in  \emph{Spitzer} IRS spectra of HR Car and G79.29+0.46; ][]{2009Umana, 2014Agliozzo}, we only fit the greybody to the MIPS24, PACS100+160, and ALMA photometry. Table \ref{tab:photometry} contains a summary of the photometry used to fit the dust. The best-fit photosphere and ionized gas free-free flux density distributions were added to the greybody for fitting the observed photometry. The free-free contribution to the ALMA measurement is significant at 0.47 to 1.01 mJy (limits of twice minimum $\chi^2$ of the free-free fit to the ATCA photometry), assuming the free-free power-law spectrum continues without turnover to the ALMA frequency. The uncertainty on the free-free contribution of $\pm0.27\,\rm mJy$ was added in quadrature with the ALMA measurement error in the fit.

A dust emissivity at $850\,\mu{\rm m}$ of $\kappa_{850}=1.7\rm\,cm^{2}\,g^{-1}$ and LMC distance of $49.97\,\rm kpc$ are assumed. 
The greybody fit yields representative parameters of $T_{\rm dust}=70.9\pm8.9\,{\rm K}$, $\beta=1.31\pm0.30$ and $M_{\rm dust}=0.055\pm0.018~M_\odot$. The reduced $\chi^2$ of the fit is 1.44. This somewhat high value appears to be mostly due to the PACS160 measurement, which we do indeed expect to be the most affected by confusion due to having the largest beam size. The greybody flux density distribution region within twice the minimum $\chi^2$ is shown as the grey band in Figure \ref{fig:sed}. The total model, including best-fit free-free, dust greybody and photosphere is shown by the black line (solid longward of $20\,\mu$m where we consider the greybody meaningful). The AKARI S11 and L15 points in Figure \ref{fig:sed} have a colour-correction applied appropriate for a $T=70\,\rm K$, $\beta=1$ greybody (correction factors are 2.712 and 2.135, respectively, AKARI IRC Data User Manual), but they still lie significantly above the photosphere and the greybody. As previously mentioned, we expect these bands, which have a wide bandwidth, to show evidence of silicate or PAH emission, like in the case of the  Galactic candidate LBV HD168625 \citep{1997Skinner,2010Umana}, or several bright mid-IR ionized gas emission lines, like in the case of HR~Car and G79.29+0.46 \citep{2009Umana, 2011UmanaB, 2014Agliozzo}, together with a possible hotter thermal component. Another source of excess at the mid-IR wavelengths could be very small grains, typically observed in HII regions and produced by erosion of large grains in the diffuse medium. Contribution from very small grains to the emissivity in the far-IR and sub-mm should be negligible \citep{2010Compiegne} and we are not aware of their observations in LBV nebulae. 

Note the departure of $\beta$ from the interstellar case ($\beta = 2$): $\beta \sim 1.3$ implies either the existence of
relatively large grains, a significant optical depth, or a range of physical dust temperatures \citep[studies of Galactic objects showed that the dust temperature decreases with increasing distance from the star, e.g.][]{1997Hutsemekers, 2017Buemi}. Large grains have been found in other LBV nebulae, such as those of AG~Car \citep{1988ApJ...329..874M,2015Vamvatira} and $\eta$~Car \citep{1986MNRAS.222..347M}, and in the ejecta of RSGs \citep[e.g.][]{2015Scicluna}. The derived temperatures are higher than typical values of interstellar dust and also dust in evolved HII regions \citep{2012Paladini}. The dust temperature is typical compared to other LBVs (e.g. \citealt{1997Hutsemekers}). Stars displaying the B[e] phenomenon have higher dust temperatures of about 500--1000~K \citep{1998A&A...340..117L}. 
A detailed modelling of the spatially integrated SED of $\eta\,\rm Car$'s Homunculus including mid- and far-IR spectroscopy by \citet{2017Morris} shows a very complex chemical composition of the dusty ejecta. Unfortunately, the current data of \rmc preclude us to derive a more detailed model. Note that \rmc has a significant fraction of ionized gas, while the Homunculus is mostly neutral, so the mid-IR spectrum might be different in the two objects.

%

\section{The nebular mass and initial mass}

\begin{table}
\centering
\caption{Summary of available LMC LBVs dust properties.}
\label{tab:LMCdust}
\begin{tabular}{lcccr}
\hline\hline 
Source & Dust Mass & Dust T & $\beta$& Ref. \\ 
& ($10^{-2}\,\rm M_\odot$)& (K) & &  \\ 
\hline
\rmc &$5.5\pm1.8$ & $62-80$ & $1.0-1.6$ & this work \\
\rmca& $0.2-2$& 71$-$90&1.5$-$2.0&[1]\\
\sixtyone &$0.5-3$& 105$-$145&0.55$-$1.5&[2]\\
RMC$\,$71 & 1& N/A& N/A&[3]\\
\hline
\multicolumn{5}{l}{\tablebib{
 [1] \citet{2017AgliozzoB}; [2] \citet{2017AgliozzoA};
 [3] \citet{2014Niyogi}.}}
\end{tabular}

\end{table}

The sub-mm emission detected with ALMA presents a different morphology than the optical and radio emission, although the contours have low statistical significance due to the low signal-to-noise ratio. The sub-mm emission on the north-east side of the nebula may be aligned with the interstellar filaments associated with \dor. An analysis of the data shows that: 1) at least two-thirds of the integrated sub-mm flux density is due to dust that extinguishes the H$\alpha$ emission in-homogeneously across the nebula; 2) dust temperatures and parameter $\beta$ values are consistent with processed material in LBV nebulae rather than with ISM dust. 

The dust mass derived in the previous Section is to date the largest value  found in an LBV nebula at sub-solar metallicities. Only a few studies on dust in the MCs LBVs have been performed, see Table \ref{tab:LMCdust}.
A study to address the contribution of LBVs to dust production at the LMC and SMC  metallicities is ongoing (Agliozzo et al., \emph{in prep}).

The gas-to-dust ratio G/D is an unknown parameter in LBVs. Because of favourable physical conditions in LBV eruptions, this parameter may be lower than in the ISM. \citet{2014Roman-Duval} report a
G/D of $\sim\!400$ in the LMC diffuse atomic ISM, although they find that in dense regions this might be lower (G/D $\sim\!200$). Assuming the lower value as representative of the \dor region and the dust mass derived in this work, we find a total nebular mass $M_{\rm TOT}$ for \rmc of $11.0\pm{3.7}~M_{\odot}$, and assuming G/D=100 (more representative of our Galaxy) $M_{\rm TOT}=5.5\pm1.8~M_{\odot}$. 

In \citet{2012Agliozzo}, we derived an approximate ionized mass of $\sim0.5-0.9~M_{\odot}$. Although the ionized mass estimate is  dependent on the assumed volume for the nebula (which is not easy to determine because of the nebula's irregular shape), it seems plausible that a large gas mass is not ionized and is located in a photo-dissociation region around the ionized nebula, mixed with the dust. The total nebular mass has an order of magnitude uncertainty, due mostly to the uncertainty of the gas to dust ratio and of the assumed dust emissivity, $\kappa_{850}$.

The stellar parameters of \rmc suggest that the star's location in the HR diagram is at the lower luminosity end of LBVs. Our derived luminosity (log$(L/L_{\odot}) = 5.32$) suggests a single-star initial mass around $25-30~M_{\odot}$. Given that the CMFGEN models suggest a (spectroscopic) current mass of $\sim 8~M_{\odot}$, it appears that \rmc has lost a large fraction of its initial mass. \rmc's position in the HR diagram is similar to that of the Yellow Hypergiant (YHG) Hen3-1379, a post-Red Supergiant (RSG) star on its way to the LBV phase, as argued by \citet{2011Lagadec} and \citet{2013Hutsemekers}. Further analysis would be needed to draw firm conclusions on whether the nebula was ejected during binary interaction, or LBV eruptions, or during a previous RSG phase. In particular, nebular abundances would need to be compared to numerical stellar evolution models of single and binary stars. \citet{2018Beasor} discussed the uncertainties of RSG mass-loss rates along the evolutionary sequence of a $16~M_{\odot}$ star, showing that such a RSG star loses a small fraction ($\sim 0.6~M_{\odot}$) of its  mass through stellar winds. However, no similar observational estimate exists for the more massive RSGs that \rmc could have evolved from.

In light of our revised low luminosity of \rmc, an interesting avenue for future study would be a re-appraisal of the evolutionary status of \rmc in comparison to its nearby environment \citep{Smith2015, Humphreys2016,Aadland2018}. \citet{Aghakhanloo17} developed models of passive dissolution of young stellar clusters, and concluded that LBV environments are inconsistent with them evolving from single stars. It would be interesting to investigate the behaviour of \rmc given its relatively low current and initial mass that we derived in this paper.


\section{The ionizing source of $\rmc$'s nebula}
The quiescent temperature of \rmc over the past $\sim 60$~yr suggests that the star is not hot enough to ionize its extended nebula. Although we cannot exclude a hotter phase for \rmc in the past, the observations suggest that an external ionizing source may be responsible for the ionization of its nebula. This was also proposed for the circumstellar nebulae of the Pistol Star \citep{1999Figer} and IRAS 18576$+$0341 \citep{2010Buemi}. 

The nearby star-forming region \dor contains $> 700$  massive stars \citep{2013Doran}, with a large fraction of them forming the well-known RMC$\,$136 cluster. \dor's giant nebula is ionized and detected at radio wavelengths (Figure \ref{fig:r143_5GHz}). \citet{1985Melnick} conclude that a large fraction of the ionization of \dor is provided by the UV photon budget produced by the early-type stars, WR stars, and blue supergiants within 25~pc of the cluster centre. \citet{2013Doran} estimate the feedback of hot luminous stars in \dor from a census of hot stars within 150~pc of RMC$\,$136 obtained by the VLT-FLAMES Tarantula Survey \citep{2011Evans}. 

We use the ionizing photon luminosities in Doran et al.\ to derive an order of magnitude estimate of the Str\"{o}mgren radius at different distances from RMC$\,$136. In particular, we use the cumulative UV photon fluxes for the sample that includes the Grand Total of stars with spectroscopic classification from their Table 9. 
We derive the Str\"{o}mgren radius as $R_{S}=(3\times S_{UV}/(4\pi n_{e}^{2}\beta_{2}))^{1/3}$, where $\beta{_2} = 3\times 10^{-13} \,\rm cm^{3}\, s^{-1}$ (for typical ISM electron temperatures of $\sim 10\,000$~K) and $S_{UV}$ is the UV photon flux per second, in the simplifying assumption that the gas is mostly hydrogen. 

Accounting for only the stars within 5~pc from the cluster centre, the integrated UV photon flux in Doran et al.\ is \hbox{$576\times10^{49}\,\rm ph\,s^{-1}$}. The Str\"{o}mgren radius decreases with the plasma density $n_e$ as $n{_e}^{-2/3}$. Thus, for two orders of magnitudes of plasma densities in the ISM and in star-forming regions, $\rm 30\, cm^{-3}$ and $\rm 300\, cm^{-3}$, $R_{S}$ varies between 56 and $12$~pc, respectively. The UV photon flux in the \dor region doubles if stars at larger distances from the cluster centre are taken into account (up to a projected distance of 150~pc). For a total of $1056\times10^{49}\,\rm ph\,s^{-1}$, $R_{S}$ varies from about 15~pc (for densities of $300\rm \, cm^{-3}$)  to 68~pc (for densities of $30\rm \, cm^{-3}$). \rmc has a projected distance of about 32~pc from RMC$\,$136. Although we are considering projected distances, it seems plausible that \rmc's nebula is ionized by an external ionizing source. 

In addition, \citet{2013Doran} estimate that about 6\% of the ionizing photons escape the region, although with large uncertainty. This implies that all hydrogen inside the region is in ionization equilibrium, including plausibly \rmc's circumstellar nebula.
A similar scenario \citep["inverted" photoionization;][]{2009Schuster} was outlined to explain the asymmetric HII region associated with the Red Supergiant NML Cyg \citep{1982Habing,1983Morris,2006Schuster}. This phenomenon is also found in the YHGs and RSGs of the Westerlund~1 cluster in our Galaxy \citep{2010Dougherty,2018Fenech,2018Andrews}.

\section{Summary and conclusions}
We report the discovery of a massive dusty component in the nebula around LBV \rmc, through resolved sub-mm observations performed with ALMA. The dust mass inferred from the available space telescope IR photometry and from the ALMA Band 7 observations is $0.055\pm0.018~M_{\odot}$ assuming $\kappa_{850}=1.7\rm\,cm^{2}\,g^{-1}$, a distance of $49.97\,\rm kpc$, and an average dust temperature of  $\sim70\,\rm K$. The representative dust temperature is derived from a modified grey-body that best fits the MIPS24, PACS100, PACS160, and ALMA 850~$\mu \rm m$ photometry (integrated over the entire nebula). An additional mid-IR excess is seen in the broad-band AKARI S11 and L15 photometry, which could be due to a combination of emission lines from the ionized gas, silicate, and PAH emission bands.

We compare the ALMA dust detection with the $8\,\rm \mu m$ \emph{Spitzer} and the H${\alpha}$ \emph{HST} images, and with the 5.5, 9, and 17 GHz ATCA maps. The morphological comparison and the extinction map suggest that the dust is distributed in a photo-dissociation region partially surrounding the ionized nebula. 

The historical light curve of \rmc and the past literature show that the previous LBV classification was likely the result of a mis-identification with another star.
However, despite the possible lack of evidence of S~Doradus cycles, the massive circumstellar nebula and the stellar parameters support the LBV classification. \rmc is at the lower luminosity end of LBVs (log$(L/L_{\odot}) = 5.32$), with an effective temperature in 2015 of $\sim 8\,500\, \rm K$.  The star may be in outburst phase for the last three decades, which would also explain the relatively high mass-loss rate of $1.0 \times 10^{-5}~M_{\odot}$~yr$^{-1}$.
The stellar wind is enhanced in He and N, and depleted in H, C, and O, confirming the evolved nature of the star.
\rmc is currently not hot enough to ionize its circumstellar nebula. While the star may have been hotter in the past, the nebula could be kept in ionization equilibrium by the UV photons escaping the 30~Dor star-forming region. 

The current stellar mass in our best-fit CMFGEN model is $\sim8~M_{\odot}$, which is very low compared to the total nebular mass of $\sim 5.5~ M_{\odot}$, derived assuming a gas-to-dust ratio of 100 and a dust mass value from the best-fit of the mid-IR to sub-mm data. This suggests that the star has already lost a large fraction of its initial mass, through past LBV eruptions or possibly binary interactions. 

Because of its brightness across the electromagnetic spectrum, \rmc is ideal for detailed studies of its chemical and dust composition and kinematics of its circumstellar nebula. Future integral field unit spectroscopy and continuum high-resolution observations (e.g. with Gemini/GMOS-S, VLT/Muse, ALMA, ELT and the \emph{James Webb Space Telescope}) will shed light on the formation of \rmc's massive circumstellar nebula.

\begin{acknowledgements}  
We thank the reviewer, Simon Clark, for their careful review of the manuscript and their constructive remarks. 
We wish to thank the staff at ALMA, ATCA and ESO who made these observations possible. We also thank Jorge Melnick for trying to find the photographic plates of his observations of \rmc.
CA acknowledges support from FONDECYT grant No. 3150463. Support to CA and GP was provided by 
the Ministry of Economy, Development, and Tourism's Millennium Science
Initiative through grant IC120009, awarded to The Millennium Institute
of Astrophysics, MAS. This paper makes use of the following ALMA data:
ADS/JAO.ALMA\#2013.1.00450.S. ALMA is a partnership of ESO
(representing its member states), NSF (USA) and NINS (Japan), together
with NRC (Canada) and NSC and ASIAA (Taiwan) and KASI (Republic of
Korea), in cooperation with the Republic of Chile. The Joint ALMA
Observatory is operated by ESO, AUI/NRAO and NAOJ. This paper also
includes data collected: at the European Organisation for Astronomical
Research in the Southern Hemisphere under ESO programme 096.D-0047(A); at the Australia Telescope Compact Array, which is part of
the Australia Telescope National Facility which is funded by the
Australian Government for operation as a National Facility managed by CSIRO. 
This work made use of PyAstronomy. This research has made use of the:  International Variable Star Index (VSX) database, operated at AAVSO, Cambridge, Massachusetts, USA; VizieR catalogue access tool, CDS,
 Strasbourg, France. The original description of the VizieR service was
 published in A\&AS 143, 23.
\end{acknowledgements}



\bibliographystyle{aa} 
\bibliography{mybib} 

\end{document}